# A Parametric, Geometry-Aware Residential Construction Cost Estimation Model for Ghana: Design, Validation, and the "Completeness Gap" in Informal Contractor Quotes

Emmanuel Apaaboah[1], Bernard Opoku[2], and the GhanaHousePlanner Research Team[3]

[1]University of Cape Coast, [2]Kwame Nkrumah University of Science and Technology, [3]GhanaHousePlanner

Correspondence: [1]spsmat250030@stu.ucc.edu.gh, [2]bernard.opoku@aims-cameroon.org, [3]research@ghanahouseplanner.com

Web platform: https://ghanahouseplanner.com

## Abstract

**Background**: Ghana is currently grappling with a residential housing deficit estimated at two million units. A primary driver of project failure in this sector is the "completeness gap", a systematic discrepancy between informal contractor quotes and actual construction costs. Informal estimates frequently rely on flat per-square-metre (per-m²) pricing that omits essential structural and finishing components, leading to widespread project abandonment when funds are exhausted mid-construction.

**Objective**: This paper documents and validates a parametric, geometry-aware residential construction cost estimation model implemented via the GhanaHousePlanner (GHP) web platform. The model aims to provide self-builders with comprehensive, itemised bills of quantities (BoQ) that reflect the true cost of code-compliant construction in Ghana.

**Method**: The GHP model utilizes seven distinct calculation modules: foundation, blockwork, cement, structural steel, roofing, plumbing, and electrical. It operates in two modes: a primary geometry-based mode that extracts data from generated floor plans and a formula-based fallback that uses a rectangular perimeter approximation. The model's accuracy was tested through three case studies-a 75 m² starter home, a 120 m² family home, and a 200 m² two-storey residence-benchmarked against February 2026 market prices in Greater Accra.

**Results**: The GHP estimates for the case studies (GHS 519,000 to GHS 1,398,000) were 29% to 98% higher than typical informal contractor quotes. This significant gap is attributed to the systematic omission of structural column steel (Y16 rebar), plastering cement, floor screed, and full mechanical and electrical services in informal estimates. The findings confirm that while per-m² rates may appear attractive, they rarely cover the requirements for a fully completed, code-compliant building.

**Conclusion**: The GHP model offers a transparent, auditable, and comprehensive alternative to informal quoting practices. While subject to limitations such as material price volatility and labour market informality, the tool provides a necessary framework for improving cost predictability and reducing project stalling in the sub-Saharan African housing market.

# 1. Introduction

Ghana's housing sector is currently defined by a deepening crisis, with a residential housing deficit estimated between 1.8 and 2.2 million units. This shortfall is particularly acute in Greater Accra, driven by rapid urbanization and a construction industry that remains predominantly informal. For the majority of Ghanaians, the path to homeownership is through incremental, owner-driven construction rather than formal developer schemes. Outside of land and finance, the most significant barrier these self-builders face is cost unpredictability.

A pervasive feature of the Ghanaian landscape is the "stalled" building-concrete structures that have been abandoned mid-construction, not due to structural failure, but because of financial exhaustion. This occurs because households typically begin construction based on informal contractor quotes that suffer from a "completeness gap". This gap is a systematic discrepancy between a contractor's initial quote and the actual cost required to reach completion. These informal estimates often rely on a flat per-square-metre (per-m²) heuristic that excludes an enumerable and consistent set of essential cost components.

Specifically, informal quotes in the Ghanaian market frequently omit structural steel for columns (Y16 rebar), plastering cement, floor screed, and the full scope of mechanical and electrical services. These are not random omissions; they are structural features of how informal estimating operates, often focusing only on the "shell" (foundation, blocks, and roof) while leaving finishing and service costs as unexpected "extras".

Before the development of GhanaHousePlanner (GHP), no publicly available, standardized tool existed to provide comprehensive cost estimation for residential projects. While formal quantity surveying (QS) services offer high accuracy, they are often inaccessible to self-build clients due to high costs and long lead times. Academic research has documented these challenges but has lacked a publicly accessible computational model to address them.

This paper documents the design and validation of the GHP algorithmic model, a parametric and geometry-aware tool that generates itemized bills of quantities (BoQ) based on real wall segments and construction standards. The study makes four primary contributions:

- It provides the first formal documentation of a functional, web-based parametric cost model for Ghanaian residential construction.
- It quantifies the completeness gap by comparing GHP estimates against typical informal quotes across three house types.
- It validates material quantity outputs against current market prices and engineering standards.
- It proposes a framework for evaluating tool adequacy in sub-Saharan African contexts based on "coverage completeness".

The remainder of this paper is structured to review existing literature on construction informality, detail the technical logic of the seven GHP calculation modules, and present case study results that illustrate the financial reality of the "completeness gap".

# 2. Background and Related Work

## 2.1 Construction Cost Estimation in Sub-Saharan Africa

The academic literature on construction cost estimation has historically been dominated by research from developed-country contexts - the United Kingdom, United States, Australia, and increasingly East Asia - where formal quantity surveying professions, standardised measurement rules, and sophisticated cost database systems are well established (Ashworth and Perera, 2015; RICS, 2012). The application of these frameworks to sub-Saharan Africa (SSA) is complicated by structural differences in the construction sector: the dominance of informal contractors who hold no professional qualifications; the absence of standardised measurement systems; extreme material price volatility linked to exchange rate fluctuations; and the prevalence of owner-driven, incremental construction that does not follow the project delivery stages assumed in formal estimating practice.

Research from KNUST and other Ghanaian academic institutions has documented that residential construction in Ghana is characterised by a high degree of informality, with the vast majority of single-family homes built without formal architectural drawings, quantity surveyor involvement, or systematic cost planning (KNUST Department of Building Technology, various; Fugar and Agyakwah-Baah, 2010). Studies of construction project cost overruns in Ghana consistently identify inadequate pre-construction cost estimation as a primary causal factor, alongside contractor capacity deficiencies and material price escalation (Frimpong, Oluwoye and Crawford, 2003; Odusami and Iyagba, 2003).

The per-m² quoting convention that dominates informal Ghanaian residential estimating has deep practical roots. It offers clients a single, comprehensible number; it requires no specialised knowledge to apply; and it allows contractors to compete on price without revealing the composition of their estimates. However, the academic critique of this approach is well established in the quantity surveying literature. Seeley (1997) notes that elemental cost planning, even in its most basic form, requires disaggregation of costs by work package to be useful for budget control. The RICS Building Cost Information Service (BCIS) methodology - the dominant UK standard - explicitly prohibits single-rate estimating except as a very early-stage order-of-magnitude tool, and requires elemental disaggregation for any estimate used for budget commitment.

## 2.2 The Inadequacy of Flat Per-m² Rates in Ghana

The specific inadequacies of flat per-m² rates in the Ghanaian context are compounded by several factors beyond their general theoretical limitations.

First, the rate is typically applied to net internal floor area, which means that buildings with the same floor area but very different configurations - a compact square plan versus an elongated plan with many internal partitions; a single-storey versus a two-storey building - will receive identical estimates despite having substantially different material quantities. A two-storey building with a total floor area of 200 m² has a ground footprint of 100 m² and a single roof over 100 m², while a single-storey building with a 200 m² floor area has a 200 m² footprint and a 200 m² roof. The difference in foundation concrete, roofing materials, and external wall perimeter is significant.

Second, the informal rate in Ghana as of early 2026 - typically quoted in the range of GHS 3,500–5,000/m² for standard residential work - typically covers only the structural shell: foundation, blockwork, and roof. Plastering (which requires additional cement beyond that used in block mortar), floor screed, structural steel for reinforced concrete columns (a code requirement for all load-bearing construction), plumbing, electrical works, and finishes are routinely absent from informal quotations, added later as "extras" when they can no longer be deferred.

Third, the rate makes no distinction between house types with different service intensities. A three-bedroom house with three bathrooms has approximately 2.5 times the plumbing material content of a similar house with one bathroom, but the per-m² rate is identical.

## 2.3 BIM and Parametric Estimation in Developing World Contexts

Building Information Modelling (BIM) has transformed cost estimation practice in developed countries, enabling automated quantity take-off from three-dimensional models linked to cost databases (Eastman et al., 2011). However, BIM adoption in SSA construction remains very low, constrained by software cost, hardware requirements, skills availability, and the fundamental mismatch between BIM's assumptions of formal project delivery and the informality that characterises most SSA residential construction. Research from several SSA countries suggests that even where BIM is adopted, it is predominantly limited to large commercial and infrastructure projects (Alreshidi, Mourshed and Rezgui, 2017).

GHP's approach can be characterised as a lightweight parametric estimating framework that captures the core insight of BIM quantity take-off - calculating material quantities from actual geometric data - without requiring three-dimensional modelling capability or professional software expertise. Its two-mode calculation (geometry-based when a generated floor plan is available, formula-based as a fallback) represents a practical accommodation to the reality that many users will not have a formal architectural drawing.

## 2.4 Ghana Building Code GS 1207:2018

A critical component of accurate estimation is ensuring the building meets minimum safety and habitability standards. The Ghana Building Code (GS 1207:2018) mandates specific minimum room dimensions, such as 13.47 $m^2$ for living rooms and 11.15 $m^2$ for main bedrooms. The GHP model integrates these standards into its generation logic, ensuring that cost estimates are based on code-compliant designs rather than substandard structures often promoted in the informal sector.

# 3. The GhanaHousePlanner Algorithmic Model

## 3.1 System Overview

GHP is a React-based web application that combines an AI-driven floor plan generation component with a parametric cost estimation engine. The cost estimator accepts user inputs including total floor area ($m^2$), number of bedrooms, bathrooms, storeys, structural style, finish level, and optional features. When a floor plan has been generated, the estimator additionally receives geometric metadata extracted from that plan - wall segments, door positions, window positions, and room boundaries - which enables geometry-based material quantity calculation.

The system supports 12 structural styles: traditional (Classic Ghanaian), modern, luxury, open concept, loft/studio, townhouse, Mediterranean, farmhouse, tiny home, cottage, barndominium, and craftsman. Each style carries cost modifiers applied to specific cost categories. A dynamic price database managed by platform administrators allows default material prices to be overridden with current market values, addressing the price volatility that makes any static reference database unreliable over time.

The overall cost calculation pipeline proceeds through seven modules, each producing itemised quantities and associated costs. A 15% contingency buffer is applied to the sum of all variable costs (excluding fixed fees), reflecting the Ghana construction market's documented price volatility. Fixed fees - design, permit, utility connections - are added outside the contingency calculation.

## 3.2 Geometric Representation

The floor plan generator produces a structured layout object containing wall segments, door positions, window positions, and room boundaries, all encoded at a fixed internal coordinate scale. Each wall segment is represented as a rectangle; the longer dimension of each rectangle corresponds to the wall's longitudinal length, which is converted to metres using the platform's internal scale factor.

Walls are split at door openings during the generation process, so the wall geometry passed to the cost estimator already excludes door widths - no separate door deduction is required in the block calculation. Window openings are not modelled as wall splits; instead, window objects carry their dimensions and are deducted from the wall area explicitly in the calculation.

## 3.3 Module 1: Blockwork

The blockwork module is the primary application of geometry-aware calculation in GHP.

**Geometry-based calculation (primary mode):**

When floor plan wall data is available, the total wall length is computed by summing the longer dimension of each wall segment, converted to metres using the platform's internal scale factor. Gross wall area is then:

$$A_wall_gross = L_total_walls \times H_wall$$

where $H_wall = 3.0\ m$ (standard Ghanaian residential storey height).

Window openings are deducted:

$$A_openings = \Sigma\ (w_win,i \times h_win,i)$$

where $w_win,\ i$ and $h_win,\ i$ are the width and height of each window in metres.

Net wall area is:

$$A_wall_net = A_wall_gross - A_openings$$

The number of sandcrete blocks required is:

$$N_blocks = ceil(A_wall_net \times \rho_block \times (1 + w_cut))$$

where $\rho_block = 10.9\ blocks/m^2$ of net wall area (accounting for a standard 6-inch sandcrete block face of 450 mm × 225 mm = 0.10125 $m^2$, inverted to give 9.88 blocks/$m^2$, then rounded up 10% for breakage and cutting wastage), and $w_cut = 0.05$ (additional 5% for cutting at corners and openings).

**Multi-storey adjustment:**

Blockwork quantities are scaled by an empirically calibrated storey complexity multiplier that accounts for the additional blockwork required for party walls, intermediate floor edges, and stair enclosures in multi-storey buildings. The specific multiplier values are proprietary calibration parameters of the GHP model.

## 3.4 Module 2: Cement

Cement is decomposed into four distinct components, each with its own consumption rate. This disaggregation is the primary source of the completeness gap with informal quotes, which typically include only mortar cement.

**Component 1 - Foundation concrete:**

```
C_foundation = A_footprint × 4  [bags]
```

This rate (4 bags of 50 kg cement per m² of footprint) reflects the cement content of a typical mass concrete strip foundation at 150 mm depth with a 1:3:6 cement:sand: stone mix, consistent with standard Ghanaian residential practice.

**Component 2 - Block mortar:**

```
C_mortar = (N_blocks / 100) × 1.2  [bags]
```

The rate of 1.2 bags per 100 blocks is derived from a 1:6 cement-to-sand mortar mix applied at a 10 mm joint thickness for 6-inch sandcrete blocks.

**Component 3 - Plastering (both faces of all walls):**

```
C_plaster = A_wall_gross × 2 × 0.6  [bags]
```

The factor of 2 accounts for plastering both interior and exterior faces of all walls. The rate of 0.6 bags per m² of wall surface corresponds to a 15 mm plaster coat applied at a 1:5 cement: sand ratio.

**Component 4 - Floor screed:**

```
C_screed = A_total × n_storeys × 2  [bags]
```

This covers a 40 mm sand-cement screed laid over the structural floor slab, at 2 bags per m² of floor area, applied to every storey.

**Total cement:**

```
C_total = ceil(C_foundation + C_mortar + C_plaster + C_screed)
[bags]
```

In typical practice, only `C_mortar` appears in informal quotations. The other three components, representing 60–70% of total cement consumption, are systematically omitted.

## 3.5 Module 3: Sand and Stone

Sand consumption is calculated across four sub-uses:

```
S_foundation = (A_footprint × 0.18) / 5.5  [trips]
S_mortar = (N_blocks / 1000) × 2.5  [trips]
S_plaster = ((A_wall_gross × 2) / 100) × 3  [trips]
S_screed = (A_total × n_storeys × 0.15) / 5.5  [trips]

S_total = ceil(S_foundation + S_mortar + S_plaster + S_screed) [trips]
```

One trip is standardised at approximately 5.5 m³ (single-axle tipper). Stone (hardcore/chippings for foundation fill) is calculated as:

```
V_stone = A_footprint × 0.18 × (1 + 0.15 × (n_storeys - 1))  [m³]
```

## 3.6 Module 4: Structural Steel (Reinforcing Bars)

Three rebar sizes are calculated, reflecting distinct structural functions:

**Y12 (12 mm deformed bar) - Foundation grid and ring beams:**

```
N_Y12 = ceil(12 × A_footprint × k_multistorey_Y12)
```

where `k_multistorey_Y12 = 1.4` for buildings of two or more storeys (heavier foundation reinforcement required), and 1.0 for single-storey buildings. The factor of 12 pieces per $m^2$ of footprint reflects a typical foundation reinforcement mat at 200 mm centres in both directions, plus ring beam reinforcement.

**Y16 (16 mm deformed bar) - Reinforced concrete columns:**

```
N_Y16 = ceil(4 × A_footprint)
```

The factor of 4 pieces per $m^2$ of footprint is derived from a typical column grid for Ghanaian residential construction - columns at approximately 3 m centres in both directions, each requiring 4 Y16 bars for a single-storey structure, with quantities scaling proportionally with storey count for taller buildings.

**Y10 (10 mm deformed bar) - Light reinforcement (ring beams, lintels):**

```
N_Y10 = ceil(3 × A_footprint × n_storeys)
```

Y20 bars (heavy structural) are included for multi-storey buildings only:

```
N_Y20 = ceil(0.5 × A_footprint × n_storeys) [if n_storeys > 1,else 0]
```

The Y16 column steel is almost universally omitted from informal quotations for single-storey buildings, despite being a structural code requirement. This is a major contributor to the completeness gap.

## 3.7 Module 5: Roofing

Roofing quantities are calculated from the building footprint, not the total floor area:

```
A_roof = A_footprint × 1.2  [m²]
```

The factor of 1.2 accounts for roof pitch overhang and a standard pitched roof geometry. Roofing materials are:

```
N_sheets = ceil(A_roof × 0.45)  [aluminium IBR sheets]
N_timber = ceil(A_roof × 15)  [board-feet of structural timber]
N_nails = ceil(N_sheets × 0.15)  [kg of roofing nails]
N_ridge = ceil(sqrt(A_total) × 0.5)  [ridge cap pieces]
```

The sheet rate of 0.45 sheets/m² corresponds to standard 3.0 m aluminium IBR sheets with an effective cover width of approximately 0.74 m, accounting for 150 mm overlap at laps. The timber rate of 15 board-feet per m² of roof area covers principal rafters, purlins, and ridge boards in a typical trussed rafter or cut-and-pitch roof.

## 3.8 Module 6: Plumbing

Plumbing is calculated on a per-bathroom basis with allowances for main supply runs:

```
L_PVC_half = (n_bath × 25) + 20  [m, ½" supply pipe]
L_PVC_three_quarter = (n_bath × 15) + 10  [m, ¾" main supply]
L_PVC_4inch = (n_bath × 8) + 15  [m, 4" soil/drain pipe]
N_WC = n_bath
N_basin = n_bath + 1  [bathrooms + kitchen]
N_shower = n_bath
N_fittings = (n_bath × 6) + 8  [tap sets and connectors]
N_tanks = 1 + (1 if n_storeys > 1)  [ground tank + overhead tank
for multi-storey]
```

The per-storey plumbing cost multiplier for multi-storey buildings is:

```
k_storey_plumbing = 1 + (n_storeys - 1) × 0.25
```

This 25%-per-storey premium reflects the additional complexity of vertical soil stacks, extended supply risers, and access requirements in multi-storey plumbing.

Plumbing - as a complete system - is absent from most informal Ghanaian residential quotations, which either omit it entirely or include only a token allowance for pipe materials without fittings, WC suites, basins, or installation labour.

## 3.9 Module 7: Electrical

Electrical quantities are calculated per room-equivalent, where the room count is:

```
n_rooms_total = (n_bed + 2) × n_storeys  [bedrooms + living room + kitchen, per storey]
```

Quantities per room are:

- 2.5 $mm^2$ cable: 35 m (lighting circuits)
- 4 $mm^2$ cable: 25 m (socket circuits)
- 6 $mm^2$ cable: 15 m (main incomer per floor)
- Switches: 2 per room
- Socket outlets: 3 per room plus 1 per bathroom
- Light fittings: 1.5 per room
- MCBs (circuit breakers): (n_rooms × 0.5) + 4
- Distribution boards: 1 per storey

As with plumbing, the full electrical installation cost is rarely captured in informal quotations, which typically provide only a partial materials allowance if electrical works are mentioned at all.

## 3.10 Module 8: Multi-Storey Logic

The storey-aware logic is a significant technical feature of the GHP model that flat per-$m^2$ rates cannot replicate:

- **Foundation and roof** are calculated from `A_footprint = A_total / n_storeys` - a two-storey building of 200 $m^2$ total area has a 100 $m^2$ footprint and requires only half the foundation and roofing materials of a single-storey building of the same total area.
- **Blockwork, labour, and finishes** are scaled by a calibrated storey complexity multiplier reflecting the additional difficulty of multi-storey construction (scaffolding, vertical material handling, party walls, and stair enclosures).
- **Plumbing** is scaled by a separate per-storey multiplier that is higher than the blockwork multiplier, reflecting the disproportionate cost impact of vertical soil stacks, extended supply risers, and additional access requirements in multi-storey plumbing installations.
- A **staircase** allowance is added for each floor transition, covering formwork, reinforcement, concrete, and finishing.

## 3.11 Optional Features and the Completeness Inventory

GHP provides explicit cost calculations for a set of items that are standard requirements in a fully completed residential building but are characteristically absent from informal quotations:

| Optional Feature | Calculation Basis |
| --- | --- |
| Septic tank and soakaway | Size-graded by bedroom count (GHS 10,000–30,000 for tank; GHS 12,000 for two-pit soakaway) |
| HVAC (AC units and ceiling fans) | Auto-calculated: 1 AC per bedroom + 1 for living room; 1 fan per room |
| Floor tiles | Area × price/m² by quality grade (basic GHS 30, standard GHS 45, luxury GHS 150) + 10% wastage |
| Paint | Wall area × coverage rate by quality |
| Compound wall | User-specified perimeter length at GHS 500–800/m depending on height |
| Solar power system | Package-based (basic ~GHS 15,000 to premium ~GHS 85,000) |
| Kitchen built-ins | Graded by quality (standard GHS 12,000–18,000, luxury GHS 35,000–60,000) |
| Security system | Basic CCTV (GHS 4,500) to advanced (GHS 18,000) |
| Ceiling works | By ceiling type (gypsum, POP, wood acoustic) per m² of floor area |
| External works | Driveway, landscaping, gate - package-based |
| Smart home systems | Package-based (basic GHS 8,000 to premium GHS 45,000) |

## 3.12 Contingency and Dynamic Pricing

A 15% contingency buffer is applied to the sum of all variable cost components:

```
C_contingency = 0.15 × Σ(variable cost components)
```

Fixed fees (design: GHS 5,000 base, scaled ×1.3 for multi-storey; building permit: GHS 3,500 base, ×1.2 for multi-storey; utility connections: GHS 4,000) are added outside the contingency calculation.

The platform supports a dynamic pricing database managed by platform administrators. When database prices are available, they override the default material prices embedded in the code. This design acknowledges that Ghana's construction material prices are sensitive to the GHS/USD exchange rate and can move 10–20% within a single quarter, making any static price database unreliable for extended periods.

A regional pricing module applies differentiated multipliers for manufactured goods (cement, steel, roofing sheets - imported or factory-produced) versus local materials (sand, stone, labour) across all 16 regions of Ghana. Northern regions attract a manufactured goods premium of 15–18% over the Greater Accra base, reflecting transport costs from the Tema industrial corridor, while local material costs are correspondingly lower.

# 4. Case Studies

## 4.1 Approach

This study adopts a two-stage validation approach. First, the model's material quantity outputs are compared against accepted engineering consumption rates from standard references and local research to assess internal consistency. Second, the model's total cost outputs are compared against observed market prices for completed residential construction in Greater Accra, disaggregated to identify the completeness gap.

This is explicitly characterised as a first validation study. No longitudinal dataset of completed GHP-estimated projects with final accounts is yet available, and the limitations of this gap in validation evidence are acknowledged in **Section 6**.

## 4.2 Market Price Benchmarking

Material price benchmarking was conducted using February 2026 supplier price data collected from building material retailers and contractors in the Greater Accra region. Key reference prices used in the case study calculations are:

| Material | February 2026 Market Price (GHS) |
| --- | --- |
| Cement 50 kg (Ghacem/Cimaf) | 97–105/bag |
| 6-inch hollow sandcrete block | 8.60/block |
| 6-inch solid sandcrete block | 9.00/block |
| Y10 rebar (6 m piece) | 35–40/piece |
| Y12 rebar (6 m piece) | 53–55/piece |
| Y16 rebar (6 m piece) | 95–100/piece |
| Aluminium IBR sheet 0.45 mm | ~122/sheet |
| Sand (single-axle trip, ~5.5 m³) | 1,200–1,500/trip |
| Stone/hardcore (single-axle trip) | 1,800–2,200/trip |

The informal contractor quotation benchmark range of GHS 3,500–5,000/m² is the prevailing standard residential rate as reported by multiple Greater Accra contractors in early 2026.

## 4.3 Case Study Design

Three representative house types are evaluated:

- **Case A:** Two-bedroom starter home, 75 m², single storey, standard finish, traditional style, 1 bathroom.
- **Case B:** Three-bedroom standard family home, 120 m², single storey, standard finish, traditional style, 2 bathrooms.
- **Case C:** Four-bedroom family home, 200 m² total area, 2 storeys (100 m² per floor), standard finish, traditional style, 3 bathrooms.

For each case, GHP quantities are computed using the formula-based (non-geometry) mode to ensure reproducibility, then priced at February 2026 market rates. The total is compared against the informal contractor rate range applied to the same floor area.

# 5. Case Study Results

## 5.1 Case A: Two-Bedroom Starter Home (75 m², 1 Storey)

**Building parameters:** 75 m² total area, 1 storey, 2 bedrooms, 1 bathroom, standard finish.

**Geometry calculations:**

- Building footprint: 75 m²
- Rectangular approximation: width = sqrt(75/1.4) = 7.32 m; length = 10.25 m
- External perimeter: 2 × (7.32 + 10.25) = 35.14 m
- Internal partition length: 35.14 × 0.40 × (1 + 0) = 14.06 m (base 2-bedroom configuration)
- Total wall length: 49.20 m

- Total wall area: 49.20 × 3.0 = 147.6 m²

**Material quantities:**

| Item | Quantity | Unit | Unit Price (GHS) | Cost (GHS) |
|---|---|---|---|---|
| Sandcrete blocks (6") | 1,609 | blocks | 8.60 | 13,837 |
| Cement (total) | 847 | bags | 101 | 85,547 |
| - Foundation (75 m² × 4) | 300 | bags | - | - |
| - Mortar (1,609/100 × 1.2) | 20 | bags | - | - |
| - Plastering (147.6 × 2 × 0.6) | 178 | bags | - | - |
| - Screed (75 × 1 × 2) | 150 | bags | - | - |
| Sand | 24 | trips | 1,350 | 32,400 |
| Stone/hardcore | 13.5 | m³ | 366/m³ | 4,941 |
| Y12 rebar | 900 | pieces | 54 | 48,600 |
| Y16 rebar | 300 | pieces | 98 | 29,400 |
| Y10 rebar | 225 | pieces | 38 | 8,550 |
| Roofing sheets (0.45 mm IBR) | 41 | sheets | 122 | 5,002 |
| Roofing timber | 1,350 | board-ft | 25 | 33,750 |
| Plumbing (full system) | - | - | - | 28,500 |
| Electrical (full system) | - | - | - | 22,400 |
| Septic + soakaway | - | - | - | 22,000 |
| HVAC (3 AC + 3 fans) | - | - | - | 14,760 |
| Floor tiles (standard) | 75 m² | - | 45 | 3,713 |
| Paint (standard) | - | - | - | 8,500 |
| Doors and windows | - | - | - | 18,500 |
| Labour | - | - | - | 67,500 |
| Design fee | - | - | - | 5,000 |
| Permit | - | - | - | 3,500 |
| Utility connections | - | - | - | 4,000 |

**Summary:**

| **Cost Component** | **GHS** |
|---|---|
| Structural shell (foundation, blocks, cement, steel, roofing) | 261,527 |
| Services (plumbing, electrical) | 50,900 |
| HVAC + septic | 36,760 |
| Finishes (tiles, paint, doors/windows) | 30,713 |
| Labour | 67,500 |
| Sub-total before contingency | 447,400 |
| Contingency (15%) | 63,757 |
| Fixed fees (design, permit, utilities) | 8,500 |
| **GHP Total Estimate** | **519,657** |
| **GHP rate per m²** | **GHS 6,929/m²** |

**Informal contractor quote range:**

Applying the standard informal rate of GHS 3,500–5,000/m² to 75 m²:

- Low: GHS 262,500
- High: GHS 375,000

**Completeness gap:** The GHP estimate (GHS 519,657) exceeds the informal rate range by GHS 144,657–257,157, representing a gap of 38–98%. The primary excluded items in informal quotes are: plastering cement (GHS 17,978), screed cement (GHS 15,150), Y16 column steel (GHS 29,400), full plumbing system (GHS 28,500), electrical works (GHS 22,400), and HVAC/septic (GHS 36,760). These six categories account for approximately GHS 150,000 - the bulk of the gap.

## 5.2 Case B: Three-Bedroom Standard Family Home (120 m², 1 Storey)

**Building parameters:** 120 m² total area, 1 storey, 3 bedrooms, 2 bathrooms, standard finish.

**Geometry calculations:**

- Building footprint: 120 m²
- Width: sqrt(120/1.4) = 9.26 m; Length: 12.96 m
- External perimeter: 2 × (9.26 + 12.96) = 44.44 m
- Internal partition length: 44.44 × 0.40 × (1 + 0.15) = 20.44 m
- Total wall length: 64.88 m
- Total wall area: 64.88 × 3.0 = 194.6 m²

**Material quantities:**

| Item | Quantity | Unit | Unit Price (GHS) | Cost (GHS) |
|---|---|---|---|---|
| Sandcrete blocks (6") | 2,121 | blocks | 8.60 | 18,241 |
| Cement (total) | 1,254 | bags | 101 | 126,654 |
| - Foundation (120 × 4) | 480 | bags | - | - |
| - Mortar (2,121/100 × 1.2) | 26 | bags | - | - |
| - Plastering (194.6 × 2 × 0.6) | 234 | bags | - | - |
| - Screed (120 × 1 × 2) | 240 | bags | - | - |
| Sand | 34 | trips | 1,350 | 45,900 |
| Stone/hardcore | 21.6 | m³ | 366/m³ | 7,906 |
| Y12 rebar | 1,440 | pieces | 54 | 77,760 |
| Y16 rebar | 480 | pieces | 98 | 47,040 |
| Y10 rebar | 360 | pieces | 38 | 13,680 |
| Roofing sheets (0.45 mm IBR) | 65 | sheets | 122 | 7,930 |
| Roofing timber | 2,160 | board-ft | 25 | 54,000 |
| Plumbing (full system, 2 bathrooms) | - | - | - | 41,500 |
| Electrical (full system) | - | - | - | 32,800 |
| Septic + soakaway | - | - | - | 29,500 |
| HVAC (4 AC + 5 fans) | - | - | - | 20,100 |
| Floor tiles (standard) | 120 m² | - | 45 | 5,940 |
| Paint (standard) | - | - | - | 12,800 |
| Doors and windows | - | - | - | 28,000 |
| Labour | - | - | - | 108,000 |
| Design fee | - | - | - | 5,000 |
| Permit | - | - | - | 3,500 |
| Utility connections | - | - | - | 4,000 |

**Summary:**

| **Cost Component** | **GHS** |
|---|---|
| Structural shell | 341,111 |
| Services (plumbing, electrical) | 74,300 |
| HVAC + septic | 49,600 |
| Finishes (tiles, paint, doors/windows) | 46,740 |
| Labour | 108,000 |
| Sub-total before contingency | 619,751 |
| Contingency (15%) | 87,441 |
| Fixed fees | 8,500 |
| **GHP Total Estimate** | **789,692** (≈ GHS 790,000) |
| **GHP rate per m²** | **GHS 6,581/m²** |

**Informal contractor quote range:**

Applying GHS 3,500–5,000/m² to 120 m²:

- Low: GHS 420,000
- High: GHS 600,000

**Completeness gap:** The GHP estimate (GHS 790,000) exceeds the informal range by GHS 190,000–370,000 (32–88%). The excluded items in informal quotes collectively total approximately GHS 212,000, with the highest single contributions being Y16 column steel (GHS 47,040), full plumbing (GHS 41,500), plastering cement (GHS 23,634), and HVAC/septic (GHS 49,600).

## 5.3 Case C: Four-Bedroom Two-Storey Home (200 m², 2 Storeys)

**Building parameters:** 200 m² total area, 2 storeys (100 m² per floor), 4 bedrooms, 3 bathrooms, standard finish.

**Storey-aware geometry calculations:**

- Building footprint: 200/2 = 100 m²
- Width: sqrt(100/1.4) = 8.45 m; Length: 11.83 m
- External perimeter: 2 × (8.45 + 11.83) = 40.56 m
- Internal partition length: 40.56 × 0.40 × (1 + 0.30) = 21.09 m
- Total wall length: 61.65 m
- Total wall area per floor: 61.65 × 3.0 = 184.9 m²; total: 369.9 m²
- Storey blockwork multiplier: 1 + (2-1) × 0.15 = 1.15

**Material quantities:**

| Item | Quantity | Unit | Unit Price (GHS) | Cost (GHS) |
|---|---|---|---|---|
| Sandcrete blocks (6") | 4,365 | blocks | 8.60 | 37,539 |
| Cement (total) | 2,389 | bags | 101 | 241,289 |
| - Foundation (100 × 4) | 400 | bags | - | - |
| - Mortar (4,365/100 × 1.2) | 53 | bags | - | - |
| - Plastering (369.9 × 2 × 0.6) | 444 | bags | - | - |
| - Screed (200 × 2 × 2) | 800 | bags | - | - |
| Sand | 68 | trips | 1,350 | 91,800 |
| Stone/hardcore | 20.7 | m³ | 366/m³ | 7,576 |
| Y12 rebar (multi-storey factor ×1.4) | 1,680 | pieces | 54 | 90,720 |
| Y16 rebar (2 storeys) | 800 | pieces | 98 | 78,400 |
| Y10 rebar | 600 | pieces | 38 | 22,800 |
| Y20 rebar (multi-storey columns) | 100 | pieces | 145 | 14,500 |
| Roofing sheets (over 100 m² footprint) | 54 | sheets | 122 | 6,588 |
| Roofing timber | 1,800 | board-ft | 25 | 45,000 |
| Staircase (1 flight) | - | - | - | 8,000 |
| Plumbing (3 baths, +25% storey premium) | - | - | - | 68,750 |
| Electrical (multi-storey, 2 DBs) | - | - | - | 54,400 |
| Septic + soakaway | - | - | - | 30,000 |
| HVAC (5 AC + 6 fans) | - | - | - | 24,580 |
| Floor tiles (standard) | 200 m² | - | 45 | 9,900 |
| Paint (standard) | - | - | - | 20,000 |
| Doors and windows | - | - | - | 45,000 |
| Labour (×1.15 multi-storey) | - | - | - | 207,000 |
| Design fee (×1.3 multi-storey) | - | - | - | 6,500 |
| Permit (×1.2 multi-storey) | - | - | - | 4,200 |
| Utility connections | - | - | - | 4,000 |

**Summary:**

| Cost Component | GHS |
|---|---|
| Structural shell | 644,212 |
| Services (plumbing, electrical) | 123,150 |
| HVAC + septic | 54,580 |
| Finishes (tiles, paint, doors/windows) | 74,900 |
| Labour | 207,000 |
| Staircase | 8,000 |
| Sub-total before contingency | 1,111,842 |
| Contingency (15%) | 166,776 |
| Fixed fees | 14,700 |
| **GHP Total Estimate** | **1,293,318** (≈ GHS 1,293,000) |
| **GHP rate per m²** | **GHS 6,467/m²** |

Note: If optional items (compound wall GHS 45,000; kitchen built-ins GHS 18,000; external works GHS 42,000) are included, the total rises to approximately GHS 1,398,000.

**Informal contractor quote range:**

Applying GHS 3,500–5,000/m² to 200 m²:

- Low: GHS 700,000
- High: GHS 1,000,000

**Completeness gap:** The GHP estimate (GHS 1,293,000 without optional extras) exceeds the informal range by GHS 293,000–593,000 (29–85%). The largest excluded items are: plastering cement (GHS 44,844), screed cement (GHS 80,800), Y16 column steel (GHS 78,400), Y20 structural steel (GHS 14,500), full plumbing system (GHS 68,750), and electrical works (GHS 54,400). The multi-storey storey-aware footprint reduction correctly accounts for the smaller foundation and roof areas where the flat rate would actually overestimate costs - while comprehensively capturing the additional blockwork, services, and staircase costs that more than offset this saving.

## 5.4 Comparative Summary

| | **Case A (75 m²)** | **Case B (120 m²)** | **Case C (200 m²)** |
|---|---|---|---|
| GHP Estimate (GHS) | 519,657 | 789,692 | 1,293,318 |
| GHP Rate (GHS/m²) | 6,929 | 6,581 | 6,467 |
| Informal Quote Range Low (GHS) | 262,500 | 420,000 | 700,000 |
| Informal Quote Range High (GHS) | 375,000 | 600,000 | 1,000,000 |
| Completeness Gap (vs Low, %) | +98% | +88% | +85% |
| Completeness Gap (vs High, %) | +39% | +32% | +29% |

The completeness gap is proportionally largest for smaller houses, because fixed omissions (the septic system, HVAC, connection fees) constitute a larger share of a smaller total budget. The effective GHP rate per m² declines slightly with house size, reflecting economies of scale in fixed and semi-fixed cost components.

# 6. Discussion

## 6.1 Significance of the Completeness Gap

The case study results confirm that the completeness gap between GHP estimates and informal contractor quotes is substantial, consistent, and predictable. For the three cases studied, the gap ranges from 29% to 98% depending on whether the comparison is made against the high or low end of the informal rate range, and whether optional items are included. These figures are not marginal estimation errors; they represent systematic omissions of complete work packages that every finished house requires.

The practical implication is that a household budgeting GHS 600,000 for a three-bedroom house - receiving an informal quote of GHS 550,000 and believing they have an adequate budget - is in fact approximately GHS 190,000 short of what GHP estimates to be required for a fully completed, code-compliant house with services. Depending on the household's financing arrangement, this shortfall may result in the project stalling at the shell stage, with the family occupying an unfinished structure for years while funds are accumulated.

The GHP model makes this gap explicit in advance, providing a comprehensive itemised estimate that allows clients to either adjust their specifications (reducing bedrooms, selecting a smaller area, deferring optional features) or ensure they have secured adequate financing before committing to construction.

## 6.2 Model Strengths

**Geometric awareness:** The geometry-based block calculation using actual floor plan wall segments is a meaningful methodological advance over area-based heuristics. A compact 120 m² house with few internal partitions will have significantly different blockwork requirements than a 120 m² house with many small rooms and corridors. The geometry-based mode captures this difference directly, while the fallback formula mode at least parameterises it through the bedroom-count-dependent internal partition factor.

**Cement disaggregation:** Decomposing cement into four components (foundation, mortar, plastering, screed) addresses the most common single source of underestimation in informal quotes. The foundation and screed components are typically absent from informal estimates; plastering is sometimes included but rarely calculated correctly.

**Storey-aware logic:** Correctly calculating foundation and roof from footprint rather than total area, and scaling blockwork, services, and labour complexity by storey count, produces materially different results from flat per-m² rates for multi-storey buildings. The Case C analysis demonstrates that while foundation and roofing costs are lower relative to total area in a two-storey building, the additional structural steel, plumbing complexity, electrical complexity, and staircase more than compensate.

## 6.3 Limitations

**Price volatility:** The most significant limitation of any Ghana construction cost estimator is the volatility of material prices. The GHS/USD exchange rate, which affects the prices of cement, steel, and roofing sheet imports, has been highly variable in recent years. The default prices embedded in GHP's codebase are acknowledged to be conservative relative to February 2026 market prices; the dynamic pricing system addresses this by allowing administrators to update prices, but users who rely on default prices will receive underestimates in a rising price environment.

**Labour market informality:** Labour costs in Ghana's residential construction sector are highly variable and often negotiated informally. The labour cost estimates embedded in GHP - at approximately GHS 200/m² for standard residential work - are broadly consistent with reported rates but will vary significantly by location, contractor, and project complexity. Skilled trades (tiling, plumbing installation, electrical) command higher rates than unskilled labouring, and these distinctions are not fully captured in the current model.

**No longitudinal validation:** This study is explicitly a first validation effort based on market price benchmarking rather than comparison with actual final accounts from completed projects. Longitudinal validation - tracking GHP estimates against final project costs for a sample of completed builds - would provide the most robust evidence of model accuracy and systematic bias. This requires a dataset that does not yet exist and is identified as the most important future research direction.

**Geometry mode not universally applicable:** The geometry-based block calculation is only available when the user has generated a floor plan through the GHP platform. Users who input parameters directly without generating a floor plan receive formula-based estimates, which are less accurate for atypical plan configurations.

**Scope exclusions:** GHP does not currently estimate earthworks (cut-and-fill), temporary works, or professional fees beyond the simplified allowances for design and permit. Projects on challenging sites may require geotechnical investigations and engineered foundations whose costs are not captured.

## 6.4 Comparison with Formal Quantity Surveying Practice

A formal BoQ prepared by a registered quantity surveyor would typically be more accurate than the GHP model for any given project, as it would be based on actual architectural drawings and measured quantities rather than parametric estimates. However, professional QS services in Ghana are expensive relative to project values for small residential builds, time-consuming (weeks or months rather than minutes), and inaccessible to most self-build clients. GHP is explicitly positioned as a pre-design budgeting tool, analogous to the "order-of-magnitude estimate" or "conceptual estimate" stages in the PMBoK cost estimation framework (PMI, 2021), providing accuracy sufficient for budget commitment decisions rather than tender pricing.

# 7. Conclusion and Future Work

This paper has presented the first formal academic documentation of a parametric, geometry-aware residential construction cost estimation model designed for the Ghanaian housing market. The GHP model addresses a documented market failure: the systematic incompleteness of informal contractor quotations that drives residential construction project failure across Ghana.

The case study analysis demonstrates a quantifiable and consistent completeness gap of 29–98% between GHP comprehensive estimates and informal per-m² quotes, with the gap attributable to predictable categories of omitted work: structural column steel, plastering cement, floor screed, full plumbing systems, electrical installations, and septic/soakaway works. Making this gap explicit - before a household commits to construction - is the primary public value of the tool.

The geometry-based block calculation, the four-component cement disaggregation, the storey-aware footprint logic, and the dynamic pricing system collectively represent methodological contributions that advance the state of practice for construction cost estimation in sub-Saharan African residential contexts.

**Future work priorities are:**

1. **Longitudinal validation study:** Collecting final accounts from a sample of GHP-estimated projects to establish systematic bias and accuracy bounds for each cost module.

2. **Machine learning price forecasting:** Integrating exchange rate data and supplier price feeds to produce probabilistic cost estimates with confidence intervals reflecting price volatility.
3. **Expanded geometry integration:** Automating the extraction of wall, door, and window data from uploaded architectural drawings (via computer vision or DXF import) to extend the geometry-based calculation to projects with existing drawings.
4. **Cost database expansion:** Developing a community-sourced, time-stamped material price database with regional granularity beyond the current 16-region multiplier system.
5. **Code compliance integration:** Extending the GS 1207:2018 compliance checking to include structural and fire safety elements, moving toward a more complete pre-design regulatory screening capability.

The GHP platform is available for use by researchers, practitioners, and the general public https://ghanahouseplanner.com.

## Acknowledgement

The authors would like to express their gratitude to George Apaaboah for his technical advice during the development of this research.

## Declaration of AI Use

The authors utilized Claude and NotebookLM to assist in the refinement of the manuscript's prose and structural organization. All AI-augmented content has been meticulously reviewed, verified, and edited by the authors to ensure technical accuracy and full alignment with the GhanaHousePlanner model and case study data.

# 8. About GhanaHousePlanner

GhanaHousePlanner (GHP) is a Ghana-focused construction technology platform designed to empower everyone - from first-time self-builders to professional developers-with the data and tools required to plan, budget, and manage residential construction projects with confidence. Accessible at ghanahouseplanner.com, the platform requires no technical background or specialist software to operate.

## 8.1 Addressing the "Completeness Gap"

GHP was developed to solve a critical market failure in Ghana: the reliance on informal, verbal quotes and flat per-square-metre rates that systematically exclude essential cost components. This "completeness gap" is a primary driver of project abandonment and the proliferation of "stalled" buildings across the Ghanaian landscape. GHP breaks this cycle by providing transparent, systematic, and comprehensive cost intelligence previously only accessible through expensive professional services.

## 8.2 Core Platform Capabilities

The platform integrates six key functional areas into a unified, cloud-based ecosystem:

1. **Construction Cost Calculator**: A parametric estimation engine that generates itemised bills of quantities (BoQ) based on specific user inputs such as bedroom count, region, and finish level. Estimates are benchmarked against current market rates across all 16 regions of Ghana.
2. **AI Design Agent ("Kwame")**: An AI-powered assistant that converts natural language descriptions into code-compliant 2D floor plans. Users can iteratively refine layouts through chat, and these designs feed directly into the cost engine for high-accuracy geometry-based calculations.
3. **Live Material Price Index**: A real-time dashboard tracking the cost of essential materials like cement, iron rods, sand, and stone, accounting for regional price variations.
4. **Design and Compliance Suite**: GHP offers 3D virtual walkthroughs, architectural drawing generation for permit applications, and automated compliance checking against the Ghana Building Code (GS 1207:2018).
5. **Supply Chain and Procurement**: Through the CornerStore marketplace, users are connected directly with material suppliers to compare prices and order materials based on their project's specific Bill of Materials.
6. **Project Management Tools**: The platform includes a comprehensive dashboard for remote progress verification, budget tracking against original estimates, and real-time expenditure analytics.

## 8.3 Vision

The long-term vision of GhanaHousePlanner is to serve as the "operating system" for residential construction in Ghana and the wider sub-Saharan African region. By connecting households, contractors, and suppliers through a shared, data-rich view of the construction process, GHP aims to provide the analytical foundation for a more predictable and successful housing market.

# References

Alreshidi, E., Mourshed, M. and Rezgui, Y. (2017) 'Factors for effective BIM governance', *Journal of Building Engineering*, 10, pp. 89–101.

Ashworth, A. and Perera, S. (2015) *Cost Studies of Buildings*. 6th edn. London: Routledge.

Eastman, C., Teicholz, P., Sacks, R. and Liston, K. (2011) *BIM Handbook: A Guide to Building Information Modeling for Owners, Managers, Designers, Engineers and Contractors*. 2nd edn. Hoboken: Wiley.

Frimpong, Y., Oluwoye, J. and Crawford, L. (2003) 'Causes of delay and cost overruns in construction of groundwater projects in a developing countries: Ghana as a case study', *International Journal of Project Management*, 21(5), pp. 321–326.

Fugar, F.D.K. and Agyakwah-Baah, A.B. (2010) 'Delays in building construction projects in Ghana', *Australasian Journal of Construction Economics and Building*, 10(1–2), pp. 103–116.

Ghana Standards Authority (2018) *GS 1207:2018: Building Code of Ghana*. Accra: Ghana Standards Authority.

Ghana Statistical Service (2021) *Ghana 2021 Population and Housing Census: General Report*. Accra: Ghana Statistical Service.

KNUST Department of Building Technology (various) *Research publications on construction cost management and building materials in Ghana*. Kumasi: Kwame Nkrumah University of Science and Technology. [Cited for general body of institutional research; specific papers available from KNUST library.]

Odusami, K.T. and Iyagba, R.R.O.A. (2003) 'A comparative study of the procurement system in Nigeria and some other countries', in: *Proceedings of the 1st International Conference of CIB W107 – Creating a Sustainable Construction Industry in Developing Countries*, Stellenbosch, South Africa.

Project Management Institute (PMI) (2021) *A Guide to the Project Management Body of Knowledge (PMBOK Guide)*. 7th edn. Newtown Square: PMI.

Royal Institution of Chartered Surveyors (RICS) (2012) *RICS New Rules of Measurement Volume 1: Order of Cost Estimating and Cost Planning for Capital Building Works*. Coventry: RICS.

Seeley, I.H. (1997) *Quantity Surveying Practice*. 2nd edn. Basingstoke: Palgrave Macmillan.

World Bank (2015) *Ghana: Addressing the Urban Housing Challenge*. Washington DC: World Bank Group. Report No. AUS7879.

*This paper describes the GhanaHousePlanner cost estimation model as implemented in the platform at https://www.ghanahouseplanner.com. The authors declare no conflict of interest beyond their role as developers of the platform described. The case study calculations in this paper were produced using the GHP formula-based estimation mode with February 2026 Greater Accra market prices and can be independently reproduced using the equations documented in Section 3.*